\documentclass[a4paper,12pt]{article}
\usepackage{amsmath,amssymb,graphicx,epstopdf,subfigure}
\usepackage[a4paper,textheight=25.7cm,textwidth=17.4cm,dvipdfm]{geometry}

\numberwithin{equation}{section}

\usepackage[colorlinks=false,pdfborder = 0 0 0,CJKbookmarks=true]{hyperref}
\usepackage{CJK}

\usepackage{tabulary}
\usepackage{longtable}

\usepackage{xcolor}
\usepackage{indentfirst}

\begin{document}
\begin{titlepage}
{}~ \hfill\vbox{ \hbox{} }\break

\rightline{USTC-ICTS-14-17}

\vskip 3 cm

\centerline{\Large \bf A Note on Instanton Effects in ABJM Theory} \vskip 0.5 cm

\renewcommand{\thefootnote}{\fnsymbol{footnote}}
\vskip 30pt \centerline{ {\large \rm  Xian-fu Wang\footnote{wangxf5@mail.ustc.edu.cn}, Xin Wang\footnote{wxin@mail.ustc.edu.cn} and Min-xin Huang\footnote{minxin@ustc.edu.cn}  } } \vskip .5cm \vskip 30pt

\begin{center}
{Interdisciplinary Center for Theoretical Study,  \\ \vskip 0.2 cm  University of Science and Technology of China,  Hefei, Anhui 230026, China}
\end{center}

\setcounter{footnote}{0}
\renewcommand{\thefootnote}{\arabic{footnote}}
\vskip 60pt
\begin{abstract}

We consider the quantum spectral problem appearing the Fermi gas formulation of the ABJM (Aharony-Bergman-Jafferis-Maldacena) matrix model. This is known to related to the refined topological string on local $\mathbb{P}^1\times \mathbb{P}^1$ Calabi-Yau geometry. In the ABJM setting the problem is formulated by an integral equation, and is somewhat different from the one formulated directly in terms of the Calabi-Yau geometry and studied in our earlier paper. We use the similar method in our earlier paper to determine the non-perturbative contributions to the quantum phase volume in the ABJM case from the Bohr-Sommerfeld quantization condition. As in our earlier paper, the non-perturbative contributions contain higher order smooth corrections beyond those required by singularity cancellations with the perturbative contributions proposed by Kallen and Marino. Our results imply possible new contributions to the grand potential of the ABJM matrix model.

\end{abstract}

\end{titlepage}
\vfill \eject


\newpage

\baselineskip=16pt

\tableofcontents

\section{Introduction}

Non-perturbative effects are usually difficult to handle in quantum physics. The problem is better understood in quantum mechanics described by a non-relativistic particle moving in a one-dimensional potential. In the pioneering works \cite{ZinnJustin:1981}, Zinn-Justin calculated the multi-instanton contributions in quantum mechanics with various forms of potential, and also proposed a generalized Bohr-Sommerfeld quantization condition which can take into account all instanton contributions. For recent expositions see \cite{ZinnJustin:2004}. The generalized Bohr-Sommerfeld quantization conditions for these models are later understood in terms of \'{E}calle's mathematical theory of resurgence \cite{Ecalle, Pham}. 

In a earlier paper \cite{Huang:201406}, we consider a class of quantum spectral problems appearing in the studies of local Calabi-Yau geometries. Here the Hamiltonians are sums of exponential functions of the quantum position and momentum operators, thus has a somewhat different form from conventional non-relativistic quantum mechanics. The all-order perturbative contributions to the quantum phase volume can be summed up by the Nekrasov-Shatashvili limit of refined topological string amplitude \cite{Nekrasov:2009, Aganagic:2011}. In the study of ABJM matrix model \cite{ABJM}, Kallen and Marino proposed a way to cancel the singularities in the perturbative quantum phase volume for certain values of the Planck constant, by using the ordinary topological string amplitude as the non-perturbative contributions \cite{Kallen:2013}. In our earlier work \cite{Huang:201406}, we generalize the idea to general toric non-compact Calabi-Yau geometries and we also discovered some more smooth non-perturbative contributions beyond those in the Kallen-Marino singularity cancellation. In a different approach, the non-perturbative effects in topological strings are studied by the use of resurgent transseries \cite{Couso-Santamaria:2014}. 

In this note we follow up on the earlier work and report some calculations regarding the ABJM matrix model, which provides the original motivation for the idea of singularity cancellation \cite{Kallen:2013}. The ABJM matrix model is important for understanding non-perturbative effects of M-theory on AdS space. The partition function and grand potential have been studied extensively in the literature \cite{Hatsuda:2012,Hatsuda:2013,Hatsuda:201211,Hatsuda:201301,Calvo:2012}, and are generalized to ABJ model \cite{Honda:2014, Kallen:2014} and more cases in \cite{Moriyama:2014, Hatsuda:2014}. It is realized that the model is closely related to the refined topological string on the local $\mathbb{P}^1\times \mathbb{P}^1$ Calabi-Yau geometry.

The classical geometry of local $\mathbb{P}^1\times \mathbb{P}^1$ model can be described by the curve
\begin{align}\label{P1P1}
e^u+z_1 e^{-u}+e^v+z_2 e^{-v}=1,
\end{align}
on $(u,v)$ plane, as in \cite{Aganagic:2011,Kallen:2013}. Here $z_1,z_2$ are the complex structure modulus parameters of the geometry. To quantize this geometry, one just need to promote the classical coordinates $u,v$ to quantum operators $\hat{u},\hat{v}$ and the curve (\ref{P1P1}) to quantum wave equation
\begin{align}\label{quantum P1}
(e^{\hat{u}}+z_1 e^{-\hat{u}}+e^{\hat{v}}+z_2 e^{-\hat{v}}-1)|\psi\rangle=0,
\end{align}
with the following convention for the commutation relation
\begin{align}
[\hat{v},\hat{u}]=\frac{i\hbar}{2},
\end{align}
and $\hbar$ is the Planck constant. In order to relate to the ABJM theory, we select the special case
\begin{align}\label{z1z2}
z_1=q^{1/2}z, \qquad z_2=q^{-1/2}z,
\end{align}
where
\begin{align}
z=e^{-2E}, \quad q=e^{\frac{i\hbar}{2}}=e^{\pi i k},
\end{align}
with $\hbar=2\pi k$. 

On the other hand, the spectral problem in the ABJM model can be formulated by an integral equation 
\begin{align}\label{ABJM}
\int_{-\infty}^\infty \rho(x_1,x_2)\phi(x_2) dx_2=e^{-E} \phi(x_1),
\end{align}
with
\begin{align}\label{kernel}
\rho(x_1,x_2)=\frac{1}{2\pi k}\frac{1}{(2\cosh \frac{x_1}{2})^{\frac{1}{2}}}\frac{1}{(2\cosh \frac{x_2}{2})^{\frac{1}{2}}}\frac{1}{2\cosh(\frac{x_1-x_2}{2k})}.
\end{align}

The two formulations  (\ref{quantum P1}) and (\ref{ABJM}) are shown to be equivalent by a transformation in \cite{Kallen:2013}. One can choose a special basis of wave functions, such that the resulting matrix of the kernel (\ref{kernel}) is a Hankel matrix \cite{Hatsuda:201211}. Due to the nice properties of the Hankel matrix, the eigenvalues can be calculated numerically to much higher precision than by using the harmonic oscillator basis in our previous paper \cite{Huang:201406}. We can also use the wave equation (\ref{quantum P1}) to compute the deformed period and the Nekrasov-Shatashvili limit, which provides the perturbative contributions to the quantum phase volume.  

In the rest of this paper, we will calculate the energy spectrum numerically and constrain the non-perturbative contributions to the quantum phase volume through Bohr-Sommerfeld quantization condition. By considering many different cases of $\hbar$, we find that there are indeed some higher order smooth non-perturbative  corrections  similar as in our earlier paper \cite{Huang:201406}.

\section{The energy spectrum from Bohr-Sommerfeld quantization condition}
The quantum deformed A-period was calculated in \cite{Aganagic:2011}, and reviewed in \cite{Huang:201406,Hatsuda:2013}. Here we also review the method for convenience. The difference equation can be easily derived by represent the equation (\ref{quantum P1}) in coordinate picture
\begin{align}
(e^u+z_1 e^{-u}-1)\psi(u)+\psi(u+\frac{i\hbar}{2})+z_2\psi(u-\frac{i\hbar}{2})=0.
\end{align}
Denoting $U=e^u, V(U)=\frac{\psi(u+\frac{i\hbar}{2})}{\psi(u)}$ and taking $z_1,z_2$ as (\ref{z1z2}), we can reformulate the difference equation as
\begin{align}
(U+\frac{\sqrt{q}z}{U}-1)+V(U)+\frac{z}{\sqrt{q}V(U/q)}=0.
\end{align}
This equation is hard to solve, while for small $z$, this equation can be solved recursively as a power series of $z$. Up to order $z^3$, the result is
\begin{align}
V(U)=1-U-\frac{q^{3/2} z}{U (q-U)}-\frac{q^5 z^2}{U (q-U)^2 \left(q^2-U\right)}+\frac{q^{19/2} z^3 \left(-q^3-q^2 U+q U^2+U\right)}{U^2 (q-U)^3 \left(q^2-U\right)^2 \left(q^3-U\right)}+\mathcal{O}(z^4).
\end{align}
The quantum A-period is
\begin{align}
\Pi_{A_I}(q^{1/2}z,q^{-1/2}z;\hbar)=\log(z_I)+\widetilde{\Pi}_{A}(q^{1/2}z,q^{-1/2}z;\hbar),
\end{align}
with $\widetilde{\Pi}_{A}(q^{1/2}z,q^{-1/2}z;\hbar)$ is given by the following residue
\begin{align}
\widetilde{\Pi}_{A}(q^{1/2}z,q^{-1/2}z;\hbar)=&-2\oint \frac{du}{2\pi i}\log (V(U))=-2\oint  \frac{dU}{2\pi i}\frac{\log (V(U))}{U}
\nonumber
\\
=&\frac{2 (q+1) z}{\sqrt{q}}+\left(5 q+\frac{5}{q}+8\right) z^2
\nonumber
\\
&+\frac{2 \left(3 q^5+31 q^4+66 q^3+66 q^2+31 q+3\right) z^3}{3 q^{5/2}}+\mathcal{O}(z^4).
\end{align}
Where $\pm$ sign corresponds to the cases $z_1,z_2$, respectively.

The perturbative and non-perturbative quantum phase volume are \cite{Kallen:2013}
\begin{align}
\text{vol}_p(E,\hbar)=&8E_{\text{eff}}^2-\frac{4\pi^2}{3}+\frac{\hbar^2}{24}+\sum_{j_L,j_R}\sum_{m,d=1}^\infty\sum_{d_1+d_2=d}\frac{\hbar d}{m}N_{j_L,j_R}^{d_1,d_2}e^{\frac{i m \hbar(d_1-d_2)}{4}}
\nonumber
\\
&\times e^{-2md E_{\text{eff}}}\frac{\sin\frac{m\hbar(2j_L+1)}{4}\sin\frac{m\hbar(2j_R+1)}{4}}{\sin^3(\frac{m\hbar}{4})},\label{pvol}
\\
\text{vol}_{np}(E,\hbar)=&\sum_{j_L,j_R}\sum_{m,d=1}^\infty\sum_{d_1+d_2=d}\frac{\hbar}{2m}N_{j_L,j_R}^{d_1,d_2}[(-1)^{md}\sin\frac{8\pi^2 m d}{\hbar}e^{-\frac{8\pi m d E_{\text{eff}}}{\hbar}}+\cdots]
\nonumber
\\
&\times \frac{(2j_R+1)\sin\frac{8\pi^2 m(2j_L+1)}{\hbar}}{\sin^2\frac{4\pi^2m}{\hbar}\sin\frac{8\pi^2 m}{\hbar}},\label{npvol}
\end{align}
where
\begin{align}
E_{\text{eff}} = -\frac{1}{2}\left(\log(z)+\widetilde{\Pi}_{A}(q^{1/2}z,q^{-1/2}z;\hbar)\right).
\end{align}
and $N_{j_L,j_R}^{d_1,d_2}$ is the refined Gopakumar-Vafa (GV) invariants with $d_1,d_2$ denoting the degrees of the two $\mathbb{P}^1$'s. 

To get high order volume, one must find the high degree of GV invariants. We can use the method in \cite{IKV}. The topological string amplitude is 
\begin{equation}
\begin{split}
 &F(\omega, t,q)=
\\&\sum_{C\in H_{2}(X,{\mathbb Z})}\sum_{n=1}^{\infty}\sum_{j_{L},j_{R}}
\frac{(-1)^{2j_{L}+2j_{R}}\,N^{(j_{L},j_{R})}_{C}\Big((t\,q)^{-nj_{L}}+\cdots+
(t\,q)^{nj_{L}}\Big)\Big((\frac{t}{q})^{-nj_{R}}+\cdots +
(\frac{t}{q})^{nj_{R}}\Big)}{n(t^{n/2}-t^{-n/2})
(q^{n/2}-q^{-n/2})}\,e^{-nT_{C}}\,.
 \end{split}
 \end{equation}
We can compute topological string amplitudes in A-model, and then compare the coefficient to get GV invariants $N^{(j_{L},j_{R})}_{C}$. However, a high degree computation will cost a lot of time. For a given degree $d=d_1+d_2$, the value of $2j_L$ or $2j_R$ has a maximum \cite{Hatsuda:2013,katz1999m,Huang:2013yta}, and $(2j_L+2j_R)$ is odd \cite{Hatsuda:2013}. This, to a great extent, reduce the number of unknown $N^{j_L,j_R}_{d_1,d_2}$. We can substitute $q,t$ with some arbitrary fractional numbers, and then solve the linear equations with respect to $N^{j_L,j_R}_{d_1,d_2}$. This method help us find the higher GV invariants up to $d=14$.
 The GV invariants up to some low degrees have already been calculated in \cite{IKV,HK:2010,katz1999m,Aganagic2004}, and we list them in table \ref{GVtable}. The higher order results are very lengthy and not be listed here. Note that the non-perturbative volume (\ref{npvol}), which stands for the instanton effects, is different from the result in \cite{Kallen:2013} with some possible new higher order corrections denoted by $\cdots$. We suspect that there may be some new higher order non-perturbative contributions, which is first uncovered in \cite{Huang:201406} for local $\mathbb{P}^2,\mathbb{F}_1$ model and another special case of local $\mathbb{P}^1\times \mathbb{P}^1$ model. In \cite{Kallen:2013}, the authors have already studied the cases $\hbar=2\pi,4\pi$, where the analytical results agree with the numerical results very well. Here, we will consider a lot of different values for $\hbar$ to fix the possible corrections.

The total quantum phase volume is then easily given by
\begin{align}
\text{vol}(E,\hbar)=\text{vol}_p(E,\hbar)+\text{vol}_{np}(E,\hbar),
\end{align}
where the singularities from perturbative and non-perturbative contributions cancel each other exactly and  the $\cdots$ should not offer new poles. This can be proved by expanding the perturbative and non-perturbative contributions around the poles and using the fact
\begin{align}
(-1)^{n(2j_L+2j_R-1)}=1,
\end{align}
from a geometric argument explained in \cite{Hatsuda:2013}. So we get a well-defined total quantum phase volume. To confirm that the possible corrections to the non-perturbative volume, we first neglect them and study the volume for some values of $\hbar$, i.e. $\hbar=6\pi,8\pi,10\pi$. At these points, the total quantum phase volume, up to first few orders, are
\begin{align}
\text{vol}(E,6\pi)=&\left(8E^2+\frac{\pi^2}{6}\right)+8\sqrt{3}\pi e^{-\frac{4E}{3}}+12\sqrt{3}\pi e^{-\frac{8E}{3}}+(32E+4)e^{-4E}-\frac{176\pi}{\sqrt{3}}e^{-\frac{16E}{3}}
\nonumber
\\
&-\frac{648\sqrt{3} \pi}{5}e^{-\frac{20E}{3}}-(208 E-1)e^{-8E}+\mathcal{O}(e^{-\frac{28E}{3}}),
\\
\text{vol}(E,8\pi)=& \left(8 E^2+\frac{4}{3}\pi ^2\right)-16 (4 \text{EE}+1) e^{-2E}+(-416 E-4) e^{-4E}+\frac{128}{9} (19-276 E)e^{-6E}
\nonumber
\\
&+\frac{37}{3} (377-3504 E) e^{-8E}+\frac{416}{75} (12197-93360 E) e^{-10E}+\mathcal{O}(e^{-12E}),
\\
\text{vol}(E,10\pi)=&\left(8 E^2+\frac{17 \pi ^2}{6}\right)-\frac{40 \left(\sqrt{10-2 \sqrt{5}} \pi \right) }{5+\sqrt{5}}e^{-\frac{4}{5}E}-\frac{20 \left(\left(\sqrt{5}-15\right) \sqrt{\frac{2}{5+\sqrt{5}}} \pi \right) }{\sqrt{5}-5}e^{-\frac{8}{5}E}
\nonumber
\\
&-\frac{80 \left(\sqrt{\frac{2}{5+\sqrt{5}}} \left(4 \sqrt{5}-25\right) \pi \right) }{3 \left(\sqrt{5}-5\right)}e^{-\frac{12}{5}E}+\mathcal{O}(e^{-\frac{16}{5}E}),
\end{align}
By using Bohr-Sommerfeld quantization condition,
\begin{align}\label{BSQ}
\text{vol}(E,\hbar)=(2n+1)\pi\hbar,
\end{align}
we can then approximately give the energy spectrum in large $E$ expansion. The zero order energy $E_0^{(n)}$ can be solved by neglecting all the exponential contributions and is given by
\begin{align}\label{E0}
E_0^{(n)}=\frac{1}{2}\sqrt{\frac{2\pi^2}{3}-\frac{\hbar^2}{48}+(n+\frac{1}{2})\pi\hbar}.
\end{align}
If $\hbar$ is not too large such that $e^{-2E_0},e^{-\frac{8\pi E_0}{\hbar}}\ll 1$, then the leading order energy is already a good approximation. In this case, we can reasonably assume that the energy spectrum can be expanded according to the form of volume as
\begin{align}
E^{(n)}(\hbar)=E^{(n)}_0 +\sum_{j,l=1}^{\infty} c_{j,l} \exp[ - 2(j+\frac{4\pi l }{\hbar} )E^{(n)}_0 ],
\end{align}
with the coefficients will be determined by using Bohr-Sommerfeld quantization condition order by order. Finally, we get the results for
$\hbar=6\pi,8\pi,10\pi$, up to the first few orders,

\begin{align}
E^{(n)}(6\pi)&=E_0-\frac{\sqrt{3} \pi }{2 E_0}e^{-\frac{4 E_0}{3}}- \left(\frac{3 \pi ^2}{8 E_0^3}+\frac{\pi ^2}{E_0^2}+\frac{3 \sqrt{3} \pi }{4 E_0}\right)e^{-\frac{8 E_0}{3}}+\mathcal{O}(e^{-4E_0}),
\\
E^{(n)}(8\pi)&=E_0+(4+\frac{1}{E_0})e^{-2E_0}-(6+\frac{31}{4E_0}+\frac{2}{E_0^2}+\frac{1}{2E_0^3})e^{-4E_0}+\mathcal{O}(e^{-6E_0}),
\\
E^{(n)}(10\pi)&=E_0+\frac{5 \left(\sqrt{10-2 \sqrt{5}} \pi \right) }{2(5+\sqrt{5})E_0}e^{-\frac{4}{5}E_0}+\mathcal{O}(e^{-\frac{8}{5}E_0}).
\end{align}
We denote this method of solving eq.(\ref{BSQ}) as BS method-1 and give the energy spectrum for the first two quantum levels $n=0,1$ in the tables \ref{BS}. Where we have taken the GV invariants to degree $d=14$. The results are up to the highest order that the volume can take for the limited degree of GV invariants. We denote these results as original results-1. What needs to be emphasized is that we have not considered the possible corrections in (\ref{npvol}) by now.

\begin{table}
  \centering
\begin{tabular}{|c|c|c|}
  \hline
  $E^{(n)}(\hbar=6\pi)$ & $n=0$ & $n=1$ \\
  \hline
  $E_0$ & \underline{2.6}82645437336248 & \underline{4.6}90521628144894 \\
  \hline
  $e^{-\frac{4E_0}{3}} $& \underline{2.65}4285425065255  & \underline{4.68940}6386064896 \\
  \hline
  $e^{-\frac{8E_0}{3}} $& \underline{2.651}873570427129  & \underline{4.6894013}78727875 \\
  \hline
  $e^{-4E_0} $& \underline{2.651}600865405600  & \underline{4.689401344}698721 \\
  \hline
  $e^{-\frac{16E_0}{3}} $& \underline{2.6515}71647386464  & \underline{4.6894013445}64906 \\
  \hline
  $e^{-\frac{20E_0}{3}} $& \underline{2.65156}8179339321  & \underline{4.6894013445}64282 \\
  \hline
  $e^{-8E_0} $& \underline{2.65156}7751543228  & \underline{4.6894013445}64280 \\
  \hline
  $e^{-\frac{28E_0}{3}} $& \underline{2.65156}7697526828  & \underline{4.6894013445}64279 \\
  \hline
  $e^{-\frac{32E_0}{3}} $& \underline{2.65156}7690592086  & same as above \\
  \hline
  $e^{-12E_0}$ & \underline{2.65156}7689692293 & same as above \\
  \hline
  $e^{-\frac{40E_0}{3}} $& \underline{2.65156}7689574768  & same as above \\
  \hline
  $e^{-\frac{44E_0}{3}} $& \underline{2.65156}7689559362  & same as above \\
  \hline
  $e^{-16E_0}$ & \underline{2.65156}7689557341 & same as above \\
  \hline
  $e^{-\frac{52E_0}{3}} $& \underline{2.65156}7689557076  & same as above \\
  \hline
  $e^{-\frac{56E_0}{3}} $& \underline{2.65156}7689557041  & same as above \\
  \hline
\end{tabular}
  \vskip 18pt
\begin{tabular}{|c|c|c|}
  \hline
  $E^{(n)}(\hbar=8\pi)$ & $n=0$ & $n=1$ \\
  \hline
  $E_0$     & \underline{2.8}67868604772738 & \underline{5.288}088419875357 \\
  \hline
  $e^{-2E_0}$ & \underline{2.881}908359824412 & \underline{5.2881953}12053630 \\
  \hline
  $e^{-4E_0}$ & \underline{2.88181}4897675003 & \underline{5.288195307144}016 \\
  \hline
  $e^{-6E_0}$ & \underline{2.8818154}32413842 & \underline{5.28819530714418}5 \\
  \hline
  $e^{-8E_0}$ & \underline{2.8818154299}17427 & same as above \\
  \hline
  $e^{-10E_0}$& \underline{2.881815429926}313 & same as above \\
  \hline
  $e^{-12E_0}$& \underline{2.881815429926297} & same as above \\
  \hline
  $e^{-14E_0}$& same as above     & same as above \\
  \hline
\end{tabular}
\vskip 18pt
\begin{tabular}{|c|c|c|}
  \hline
  $E^{(n)}(\hbar=10\pi)$ & $n=0$ & $n=1$ \\
  \hline
  $E_0$ & 2.973469456595984 & \underline{5.7}89260022838205 \\
  \hline
  $e^{-\frac{8}{5}E_0} $ & \underline{3.07}4405209050261 & \underline{5.79369}3808223199 \\
  \hline
  $e^{-\frac{16}{5}E_0}$ & \underline{3.072}309476558468 & \underline{5.7936946}88558851 \\
  \hline
  $e^{-\frac{24}{5}E_0}$ & \underline{3.0724}09334483011 & \underline{5.7936946}87552826 \\
  \hline
  $e^{-\frac{32}{5}E_0}$ & \underline{3.0724}03095441525 & \underline{5.7936946}87553392 \\
  \hline
  $e^{-8E_0}$ & \underline{3.0724}03458426817 & same as above \\
  \hline
  $e^{-\frac{48}{5}E_0}$ & \underline{3.0724}03440986847 & same as above \\
  \hline
  $e^{-\frac{56}{5}E_0}$ & \underline{3.0724}03441187647 & same as above \\
  \hline
\end{tabular}
  \caption{The energy $E^{(n)}$ from the large $E$ expansion by using Bohr-Sommerfeld quantization condition (BS method-1), for the first two quantum levels $n=0,1$, for the cases of $\hbar=6\pi,8\pi,10\pi$. Here we have taken the GV invariants to degree $d=14$. The results are up to the highest order that the volume can be given for the limited degree of GV invariants. We denote these results as original results-1, which means that the possible corrections in (\ref{npvol}) have not considered by now. We underline the digits that match with numerical results.}\label{BS}
\end{table}

If $\hbar$ is large such that the right hand side of (\ref{E0}) tends to zero, this BS method-1 would break down for the not too small $e^{-E_0}$. Especially when $\hbar>4(\sqrt{36 n^2+36 n+11}+6 n+3)\pi$, the zero order energy wouldn't exist. Then the leading order energy would be given by including the lowest order of exponential contribution and the equation (\ref{BSQ}) can only be solved numerically. In this case, we expand the volume up to some order based on the degree of the GV invariants, and then directly approximately solve the equation (\ref{BSQ}) by using the language `FindRoot' of Wolfram mathematica software. For the sake of convenience, we denote this kind of numerical method as BS method-2. We list the correspondent solutions of eq.(\ref{BSQ}) order by order in table \ref{BS2} to compare with the results of BS method-1. Here we also take the GV invariants to degree $d=14$ and the results are up to the highest order that the volume can be given for the limited degree of GV invariants. We denote these results as original results-2 and we still have not considered the possible corrections in (\ref{npvol}).

Now, we have already obtained the energy spectrum of local $\mathbb{P}^1\times\mathbb{P}^1$ model for $\hbar=6\pi,8\pi,10\pi$, through solving the Bohr-Sommerfeld quantization condition (\ref{BSQ}). In the next part, we will give an infinite dimensional Hankel matrix $M$, and the corresponding spectrum problem is equivalent to the integral equation (\ref{ABJM}). We will take its solutions as the numerical results, and by comparing them with the original results-1,2, we can confirm whether the structure of the quantum phase volumes (\ref{pvol})(\ref{npvol}) is correct.

\begin{table}
  \centering
\begin{tabular}{|c|c|c|}
  \hline
  $E^{(n)}(\hbar=6\pi)$ & $n=0$ & $n=1$ \\
  \hline
  $E_0$ & \underline{2.6}82645437336248 & \underline{4.6}90521628144894 \\
  \hline
  $e^{-\frac{4E}{3}} $& \underline{2.65}2977020840839 & \underline{4.68940}4590794601 \\
  \hline
  $e^{-\frac{8E}{3}} $& \underline{2.651}615846308570 & \underline{4.6894013}59187324 \\
  \hline
  $e^{-4E} $& \underline{2.65156}1640192892 & \underline{4.6894013445}05450 \\
  \hline
  $e^{-\frac{16E}{3}} $& \underline{2.65156}7317715063 & \underline{4.6894013445}64030 \\
  \hline
  $e^{-\frac{20E}{3}} $& \underline{2.65156}7683261769 & \underline{4.6894013445}64279 \\
  \hline
  $e^{-8E} $& \underline{2.65156}7691609416 & same as above \\
  \hline
  $e^{-\frac{28E}{3}} $& \underline{2.65156}7689691920 & same as above \\
  \hline
  $e^{-\frac{32E}{3}} $& \underline{2.65156}7689558157 & same as above \\
  \hline
  $e^{-12E}$ & \underline{2.65156}7689556235 & same as above \\
  \hline
  $e^{-\frac{40E}{3}} $& \underline{2.65156}7689556982 & same as above \\
  \hline
  $e^{-\frac{44E}{3}} $& \underline{2.65156}7689557036 & same as above \\
  \hline
  $e^{-16E}$ & same as above & same as above \\
  \hline
  $e^{-\frac{52E}{3}} $& same as above & same as above \\
  \hline
  $e^{-\frac{56E}{3}} $& same as above & same as above \\
  \hline
\end{tabular}
  \vskip 18pt
  \begin{tabular}{|c|c|c|}
  \hline
  $E^{(n)}(\hbar=8\pi)$ & $n=0$ & $n=1$ \\
  \hline
  $E_0$     & \underline{2.8}67868604772738 & \underline{5.288}088419875357 \\
  \hline
  $e^{-2E}$ & \underline{2.881}556529844591 & \underline{5.288195}290191258 \\
  \hline
  $e^{-4E}$ & \underline{2.8818}07934722780 & \underline{5.28819530714}0165 \\
  \hline
  $e^{-6E}$ & \underline{2.881815}173576993 & \underline{5.28819530714418}4 \\
  \hline
  $e^{-8E}$ & \underline{2.88181542}0340901 & \underline{5.28819530714418}5 \\
  \hline
  $e^{-10E}$& \underline{2.881815429}547348 & same as above \\
  \hline
  $e^{-12E}$& \underline{2.8818154299}10733 & same as above \\
  \hline
  $e^{-14E}$& \underline{2.88181542992}5640 & same as above \\
  \hline
\end{tabular}
\vskip 18pt
\begin{tabular}{|c|c|c|}
  \hline
  $E^{(n)}(\hbar=10\pi)$ & $n=0$ & $n=1$ \\
  \hline
  $E_0$ & 2.973469456595984 & 5.789260022838205 \\
  \hline
  $e^{-\frac{8}{5}E} $ & \underline{3.0}69212559513221 & \underline{5.79369}2151779575 \\
  \hline
  $e^{-\frac{16}{5}E}$ & \underline{3.0724}21826221524 & \underline{5.7936946}87737750 \\
  \hline
  $e^{-\frac{24}{5}E}$ & \underline{3.0724}06156802973 & \underline{5.7936946}87553717 \\
  \hline
  $e^{-\frac{32}{5}E}$ & \underline{3.0724}03481354253 & \underline{5.7936946}87553392 \\
  \hline
  $e^{-\frac{40}{5}E}$ & \underline{3.0724}03439242124 & same as above \\
  \hline
  $e^{-\frac{48}{5}E}$ & \underline{3.0724}03441187597 & same as above \\
  \hline
  $e^{-\frac{56}{5}E}$ & \underline{3.0724}03441284761 & same as above \\
  \hline
\end{tabular}
  \caption{The energy $E^{(n)}$ from the large $E$ expansion by using Bohr-Sommerfeld quantization condition (BS method-2),
 for the first two quantum levels $n=0,1$, for the cases of $\hbar=6\pi,8\pi,10\pi$. Here we have taken the GV invariants to degree $d=14$. The results are up to the highest order that the volume can be given for the limited degree of GV invariants. We denote these results as original results-2, which means that the possible corrections in (\ref{npvol}) have not considered by now. We underline the digits that match with numerical results.}\label{BS2}
\end{table}

\section{Numerical results and corrections}
The eigenvalue equation (\ref{ABJM}) in ABJM theory, which is equivalent to the local $\mathbb{P}^1\times \mathbb{P}^1$ model (\ref{quantum P1}), as we have already mentioned, is also equivalent to the eigenvalue equation for an infinite dimensional Hankel matrix $M$ with the matrix elements given by \cite{Hatsuda:2012}
\begin{align}\label{Hankel}
M_{nm}=\frac{1}{8\pi k}\int_{-\infty}^\infty dq \frac{\tanh^{n+m}(\frac{q}{2k})}{\cosh(\frac{q}{2})\cosh^2(\frac{q}{2k})}=\frac{1}{4\pi}\int_{-1}^{1}dt\frac{t^{n+m}}{T_k(1/\sqrt{1-t^2})},
\end{align}
where $T_k(x)$ is the $k$-th Chebyshev polynomial of the first kind. Note that if $n+m$ is odd, $M_{nm}$ will be zero because of the odd integrand.

For $k=1,2$ cases, $M_{nm}$ has exact expression and can be found in \cite{Hatsuda:2012}. For $k=1$,
\begin{align}
M_{nm}^{(k=1)}=\frac{C_{\frac{n+m}{2}}}{2^{n+m+3}},
\end{align}
where $C_n$ is the Catalan number
\begin{align}
C_n=\frac{(2n)!}{(n+1)!n!}.
\end{align}
For $k=2$,
\begin{align}
M_{nm}^{(k=2)}=\frac{1}{4\pi}\left[-\frac{2}{n+m+1}+\psi\left(\frac{n+m+3}{4}\right)-\psi\left(\frac{n+m+1}{4}\right)\right],
\end{align}
where $\psi(x)=\Gamma'(x)/\Gamma(x)$ is the digamma function.

Through solving the eigenvalues of these two Hankel matrices, we can compare the energy spectrum of the integral equation (\ref{ABJM}) in ABJM theory and the energy spectrum from local $\mathbb{P}^1\times \mathbb{P}^1$ model (\ref{quantum P1}) that is computed by using Bohr-Sommerfeld quantization condition for $k=1,2$. It turns that they agree with each other very well and can be found in \cite{Kallen:2013}.

\begin{table}
  \centering
  \begin{tabular}{|c|c|c|}
  \hline
  $E^{(n)}(\hbar=6\pi)$ & $n=0$ & $n=1$ \\
  \hline
  $2500\times 2500$ & 2.651568337168878 & 4.689401344572731 \\
  \hline
  $3000\times 3000$ & \underline{2.6515683371688}68 & 4.689401344571483 \\
  \hline
  $3500\times 3500$ & \underline{2.65156833716886}3 & \underline{4.68940134457}0947 \\
  \hline
  $4000\times 4000$ & \underline{2.65156833716886}1 & \underline{4.689401344570}687 \\
  \hline
  $4500\times 4500$ & \underline{2.65156833716886}0 & \underline{4.689401344570}548 \\
  \hline
  \end{tabular}
  \vskip 18pt
  \begin{tabular}{|c|c|c|}
  \hline
  $E^{(n)}(\hbar=8\pi)$ & $n=0$ & $n=1$ \\
  \hline
  $1000\times 1000$ & 2.881815429926298 & 5.288195307144391 \\
  \hline
  $1500\times 1500$ & \underline{2.88181542992629}7 & \underline{5.288195307144}213 \\
  \hline
  $2000\times 2000$ & \underline{2.881815429926297} & \underline{5.288195307144}192 \\
  \hline
  $2500\times 2500$ & same as above     & \underline{5.2881953071441}88 \\
  \hline
  \end{tabular}
  \vskip 18pt
  \begin{tabular}{|c|c|c|}
  \hline
  $E^{(n)}(\hbar=10\pi)$ & $n=0$ & $n=1$ \\
  \hline
  $1000\times 1000$ & 3.07243583602644632248 & 5.79369469107338212259 \\
  \hline
  $1500\times 1500$ & \underline{3.07243583602644632}103 & \underline{5.79369469107338}163169 \\
  \hline
  $2000\times 2000$ & \underline{3.07243583602644632}092 & \underline{5.793694691073381}59262 \\
  \hline
  $2500\times 2500$ & \underline{3.0724358360264463209}0 & \underline{5.7936946910733815}8636 \\
  \hline
  \end{tabular}
  \caption{The energy $E^{(n)}$ from the Hankel matrix, for the first two quantum levels $n=0,1$, for the cases of $\hbar=6\pi,8\pi,10\pi$. The leftmost column stand for the dimensions of Hankel matrix. These finite dimensions can already approximately give the eigenvalues of the integral equation (\ref{ABJM}). We underline the digits that is same with last row.}\label{numerical}
\end{table}

It is hard to get $M_{nm}$ exactly when $k$ takes other values. So we compute the integral (\ref{Hankel}) numerically, and take high dimensional Hankel matrix to find the approximate eigenvalues. The results for $k=3,4,5$ or $\hbar=6\pi,8\pi,10\pi$ are listed in table \ref{numerical}. From these tables, we find that the energy of the ground state converges better than the first excited state. We also find that the eigenvalues of Hankel matrix converge very fast for large $k$. For $k=5(\hbar=10\pi)$ as example, the convergence is already very good when the matrix dimension is 1000. While the most important discovery is that, by comparing tables \ref{BS} with tables \ref{numerical}, or tables \ref{BS2} with tables \ref{numerical}, the energy spectrum of the integral equation (\ref{ABJM}) does not match well with the Bohr-Sommerfeld quantization method for $k=3,5(\hbar=6\pi,10\pi)$, although the energy spectrum of $k=4(\hbar=8\pi)$ still match very well in the two sides. This means that the quantum phase volume needs to be revised and there exist higher order corrections.

To find the revised phase volume, We can take different values of $\hbar$ and compare the correspondent results to find the higher order corrections. In \cite{Huang:201406}, the authors guessed some corrections to the non-perturbative phase volume for local $\mathbb{P}^2,\mathbb{F}_1$ model and another special case of local $\mathbb{P}^1\times \mathbb{P}^1$ model. At there, the authors took $z_1=z_2=z=e^{-2E}$ for local $\mathbb{P}^1\times \mathbb{P}^1$ model. In this paper, we have taken the values of $z_1,z_2$ as (\ref{z1z2}). However, this could only change the perturbative parts of the volumes of the two cases. The non-perturbative parts should be same. So we directly use the results in \cite{Huang:201406} and give the modified non-perturbative quantum phase volume, after some changes of sign for the different conventions, as
\begin{align}\label{corrections}
\text{vol}_{np}(E,\hbar)=&\sum_{j_L,j_R}\sum_{m,d=1}^\infty\sum_{d_1+d_2=d}\frac{\hbar}{2m}N_{j_L,j_R}^{d_1,d_2}\frac{(2j_R+1)\sin\frac{8\pi^2 m(2j_L+1)}{\hbar}}{\sin^2\frac{4\pi^2m}{\hbar}\sin\frac{8\pi^2 m}{\hbar}}
\nonumber
\\
&\times [\sum_{j=1}^\infty c_j(\frac{2\pi^2 m d}{\hbar})(-1)^{jmd}e^{-\frac{8j\pi m d E_{\text{eff}}}{\hbar}}], \qquad \text{with the following coefficients}
\nonumber
\\
c_1(x)&=\sin(4x), \qquad c_2(x)=c_3(x)=0,
\nonumber
\\
c_4(x)&=\sin^2(2x)\sin(16x), \quad c_5(x)=4\sin^2(4x)\sin(20x),
\nonumber
\\
c_6(x)&=8\left[3 \sin^2(4 x) \sin^2(6 x)+ \sin^2(2 x) \sin^2(8 x)+ \sin^2(10 x)\right] \sin(24 x)
\nonumber
\\
&\quad\cdots.
\end{align}

\begin{table}
  \centering
  \begin{tabular}{|c|c|c|}
    \hline
    &$\hbar=6\pi$,n=0 & $\hbar=6\pi$,n=1 \\
    \hline
    Original results-1 & 2.651567689557041  & 4.689401344564279  \\
    \hline
    Original results-2 & 2.651567689557036  & 4.689401344564279  \\
    \hline
    Revised results-1 & 2.651568337794482  & 4.689401344570317  \\
    \hline
    Revised results-2 & 2.651568337794482  & 4.689401344570317  \\
    \hline
    Numerical results & 2.651568337168860 & 4.689401344570548 \\
    \hline
  \end{tabular}
  \vskip 18pt
   \begin{tabular}{|c|c|c|}
    \hline
    &$\hbar=10\pi$,n=0 & $\hbar=10\pi$,n=1 \\
    \hline
    Original results-1 & 3.072403441187647  & 5.793694687553392  \\
    \hline
    Original results-2 & 3.072403441284761  & 5.793694687553392  \\
    \hline
    Revised results-1 & 3.072437271348499  & 5.793694691073543  \\
    \hline
    Revised results-2 & 3.072437272189056  & 5.793694691073543  \\
    \hline
    Numerical results & 3.072435836026446  & 5.793694691073382 \\
    \hline
  \end{tabular}
  \caption{The revised energy spectrum. The leftmost column labels the different methods of getting the energy spectrum. The GV invariants are still taken to degree 14 and all these results are gotten from the highest order volume that given by the finite degree GV invariants.}\label{revised-1}
\end{table}
It can easily be proved that there is no singularity in the higher order corrections, as the self-consistency requires. In the below, we will use this up to sixth order corrected non-perturbative phase volume to calculate the revised energy spectrum. We first calculate the revised energy spectrum for $k=3,5(\hbar=6\pi,10\pi)$. The method is similar as before and the results are given in table \ref{revised-1}, denoted as revised results-1,2, corresponding to the  results solved from BS method-1,2 respectively. Here the GV invariants also are taken to degree 14 and the results are up to the highest order that the volume can be given for the limited degree of GV invariants. We can find the revised results-1,2 match better with numerical results than the original results-1,2, although the ground state energy of $k=5$ case is only refined a little. Note that the corrections will not change the $k=4(\hbar=8\pi)$ energy spectrum, which can be found directly from (\ref{corrections}) because of the vanishing of $c_4,c_5,c_6$ corrected terms for $k=4(\hbar=8\pi)$. Of course, this is a necessary condition, since the original results-1,2 of $k=4(\hbar=8\pi)$ case agree with numerical results very well. For $k$ takes other integral values, i.e. $k=6,7,8,9$, the convergence of the ground state energy is so bad that we will not consider it. We give the first excited energy spectrum for these cases in table \ref{revised-3}. We can find that the revised energy spectrum are also better than the original results in various degree.
\begin{table}
  \centering
  \begin{tabular}{|c|c|c|c|c|}
    \hline
    &$\hbar=12\pi$,n=1 & $\hbar=14\pi$,n=1&$\hbar=16\pi$,n=1 & $\hbar=18\pi$,n=1 \\
    \hline
    Original results-1 & 6.231420022040951   & 6.617363191115401  & 6.962333123855871 & 7.274031778026054  \\
    \hline
    Original results-2 &  6.231420022040951  &  6.617363191115401 &6.962333123855872 & 7.274031778026025 \\
    \hline
    Revised results-1 &  6.231419980183149   & 6.617363034760174  &6.962333317708223 &7.274033633710666 \\
    \hline
    Revised results-2 &  6.231419980183149  & 6.617363034760174  &6.962333317708223 & 7.274033633710621 \\
    \hline
    Numerical results &  6.231419980189533  & 6.617363034746669 & 6.962333315166653 & 7.274033628824735\\
    \hline
  \end{tabular}
\caption{The revised first excited energy spectrum. The leftmost column labels the different methods of getting the energy spectrum. The GV invariants are still taken to degree 14 and all these results are gotten from the highest order volume that given by the finite degree GV invariants. All the numerical results come from $2500\times 2500$ Hankel matrix.}\label{revised-3}
\end{table}

In order to precisely check the corrections, we need different cases of $\hbar$ or $k$. If $\hbar$ is too small, then $E_0$ is also too small. The non-perturbative contributions are very small and can be ignored. Besides, the volume will converge slowly for small $e^{-E_0}$ or $e^{-E}$ such that we have to calculate the GV invariants to high degrees and the eigenvalues for Hankel matrix also converge too slowly which force us to take very high matrix dimensions. While if $\hbar$ is very large, we also need very high GV invariants for non-perturbative part. Because we have to take the same order for perturbative part and non-perturbative part to cancel singularities. Finally, we find the best range is $6\pi\leqslant \hbar \leqslant 10\pi$.\footnote{In these cases, $k$  are fractional numbers. Mathematically, the group $U(N)_k$ with fractional number $k$ is not meaningless.} Taking the corrections (\ref{corrections}) into account, we give the revised results for some cases of $\hbar$ in tables \ref{revised-2}. Here, we still take the GV invariants to degree 14 and the results are up to the highest order that the volume can be given for the limited degree of GV invariants. From these tables, we can find that the revised results are close to the numerical results much more. So the assumption (\ref{corrections}) is reasonable.

In \cite{Kallen:2013}, the authors did not consider the corrections (\ref{corrections}) and gave the energy spectrum for $k=1,2(\hbar=2\pi,4\pi)$, which seems have no contradictions. While we find here that the reasons for the disappearance of the corrections for  $k=1,2(\hbar=2\pi,4\pi)$ is not that there are no corrections, but that the corrections happen to be zero for $k=1,2(\hbar=2\pi,4\pi)$ and can easily be found in (\ref{corrections}). If we think over other cases $k$ or $\hbar$, like the cases we take in this paper, we find that we have to take the corrections into account. Otherwise, the energy spectrum of the integral equation (\ref{ABJM}) in ABJM theory will not match the energy spectrum from local $\mathbb{P}^1\times \mathbb{P}^1$ model (\ref{quantum P1}) computed by using Bohr-Sommerfeld quantization condition.

\begin{table}
  \centering
  \begin{tabular}{|c|c|c|}
    \hline
    &$\hbar=\frac{20\pi}{3}$,n=0 & $\hbar=\frac{20\pi}{3}$,n=1 \\
    \hline
    Original results-1 & 2.733676534406995  & 4.901574439599164  \\
    \hline
    Original results-2 & 2.733676534404727  & 4.901574439599164   \\
    \hline
    Revised results-1 &  2.733675371844009  & 4.901574439580131  \\
    \hline
    Revised results-2 & 2.733675371841705  & 4.901574439580131  \\
    \hline
    Numerical results & 2.733675375316184   & 4.901574439580357 \\
    \hline
  \end{tabular}
 \vskip 18pt
  \begin{tabular}{|c|c|c|}
    \hline
    &$\hbar=7\pi$,n=0 & $\hbar=7\pi$,n=1 \\
    \hline
    Original results-1 & 2.772592538059461  & 5.002639669369460  \\
    \hline
    Original results-2 & 2.772592538079125  & 5.002639669369460   \\
    \hline
    Revised results-1 & 2.772589280709867   & 5.002639669306602   \\
    \hline
    Revised results-2 & 2.772589280729921  & 5.002639669306602  \\
    \hline
    Numerical results & 2.772589281190162 & 5.002639669306633 \\
    \hline
  \end{tabular}
  \vskip 18pt
   \begin{tabular}{|c|c|c|}
    \hline
    &$\hbar=\frac{22\pi}{3}$,n=0 & $\hbar=\frac{22\pi}{3}$,n=1 \\
    \hline
    Original results-1 & 2.810202803836068  & 5.100642879112707  \\
    \hline
    Original results-2 & 2.810202803909060  &  5.100642879112707  \\
    \hline
    Revised results-1  & 2.810198095189728  & 5.100642879001701  \\
    \hline
    Revised results-2 & 2.810198095262721  & 5.100642879001701  \\
    \hline
    Numerical results & 2.810198088750750  & 5.100642879001723 \\
    \hline
  \end{tabular}
  \vskip 18pt
   \begin{tabular}{|c|c|c|}
    \hline
    &$\hbar=\frac{15\pi}{2}$,n=0 & $\hbar=\frac{15\pi}{2}$,n=1 \\
    \hline
    Original results-1 & 2.828543258509623  & 5.148555542445873  \\
    \hline
    Original results-2 & 2.828543258510403   & 5.148555542445873   \\
    \hline
    Revised results-1  & 2.828538546745810  & 5.148555542323129   \\
    \hline
    Revised results-2 & 2.828538546746600  &  5.148555542323129 \\
    \hline
    Numerical results & 2.828538539417836  & 5.148555542323141 \\
    \hline
  \end{tabular}
  \vskip 18pt
  \begin{tabular}{|c|c|c|}
    \hline
    &$\hbar=\frac{17\pi}{2}$,n=0 & $\hbar=\frac{17\pi}{2}$,n=1 \\
    \hline
    Original results-1 & 2.932639124280282  & 5.422092462549633  \\
    \hline
    Original results-2 & 2.932639116891402   & 5.422092462549633   \\
    \hline
    Revised results-1  & 2.932651791472953   & 5.422092463115862    \\
    \hline
    Revised results-2 & 2.932651784292468   & 5.422092463115862   \\
    \hline
    Numerical results & 2.932651829154347  &  5.422092463115864 \\
    \hline
  \end{tabular}
  \vskip 18pt
   \begin{tabular}{|c|c|c|}
    \hline
    &$\hbar=9\pi$,n=0 & $\hbar=9\pi$,n=1 \\
    \hline
    Original results-1 & 2.981221313806712  & 5.550711291040934  \\
    \hline
    Original results-2 & 2.981221313497732   & 5.550711291040934   \\
    \hline
    Revised results-1  & 2.981251022704565   & 5.550711292727979    \\
    \hline
    Revised results-2 &  2.981251022446365  &  5.550711292727979  \\
    \hline
    Numerical results & 2.981251135625010  & 5.550711292727985 \\
    \hline
  \end{tabular}
  \caption{The revised energy spectrum. The leftmost column labels the different methods of getting the energy spectrum. The GV invariants are still taken to degree 14 and all these results are gotten from the highest order volume that given by the finite degree GV invariants. All the numerical results come from $2500\times 2500$ Hankel matrix.}\label{revised-2}
\end{table}

\section{Conclusion} \label{conclusion}

We discuss the implications of our results. The grand potential of the ABJM theory has been studied extensively in the literature, in e.g.  \cite{Marino:2011,Hatsuda:2012,Hatsuda:2013,Kallen:2013,Hatsuda:201211,Hatsuda:201301,Drukker:2010,Calvo:2012,Honda:2014}, and can be expressed in terms of the quantum spectrum of the ABJM theory defined by the integral equation (\ref{ABJM}) as  
\begin{align}
J(\mu,k)=\sum_{n\geqslant 0}\log (1+e^{\mu-E_n}). 
\end{align}
The grand potential consists of the perturbative contributions, worldsheet instanton and membrane instanton contributions. In particular, the worldsheet instanton contributions are of integer powers of $e^{-\frac{\mu}{k}}$ and can be derived from the strong coupling limit of the t'Hooft expansion of the ABJM matrix model. 

The connection with the local $\mathbb{P}^1\times \mathbb{P}^1$ geometry first appeared in the work \cite{AKMV}, where it is shown that the weak coupling t'Hooft expansion of the Chern-Simons matrix model on lens space is equivalent to the expansion of topological string amplitudes near an orbifold point. Here the Chern-Simons matrix model on lens space is basically equivalent to the ABJM matrix model by an analytic continuation of the rank of the matrix to negative value. One can solve the higher genus topological string amplitudes using the B-model method of holomorphic anomaly equation and boundary conditions near special points, see e.g. \cite{HKR}. With the exact higher genus formulae for free energy and partition function, one can then make a strong coupling expansion for the t'Hooft coupling constant \cite{Drukker:2010}. It was then shown that the integral transformation from the partition function to the grand potential gives rise to the usual large volume expansion of the topological free energy in terms of Gopakumar-Vafa invariants \cite{Hatsuda:201211}. 
 
The grand potential is related to the quantum phase volume studied in this note through a Mellin transform \cite{Kallen:2013}. If our calculations are correct, then the world-sheet instanton contributions to the grand potential has more corrections than those from the ordinary topological free energy. Since our extra contributions are non-singular and first appear at the 4th order, the implied correction to the strong t'Hooft coupling expansion of the matrix model free energy should appear first at genus one and the 4th instanton. On the other hand, we find no problem in the beautiful arguments and calculations \cite{Drukker:2010, Hatsuda:201211, Kallen:2013} which lead to the proposal of non-perturbative  phase volume for the quantum spectral problem (\ref{ABJM}). In particular, the checks between the worldsheet instanton contributions for the ABJM grand potential and the strong t'Hooft coupling expansion of the matrix model free energy were performed to at least genus one and the 4th instanton in \cite{Hatsuda:201211}. Thus our extra contributions seems quite puzzling. 

A possible explanation of our extra contributions may come from some subtleties of the extrapolation from weak to strong t'Hooft coupling in the matrix model. Here the B-model higher genus formulae are valid for the complex structure parameters over the entire complex plane. However, the matrix model t'Hooft coupling constant is only a local flat coordinate for expansion near the orbifold point of the $\mathbb{P}^1\times \mathbb{P}^1$ geometry, defined by the corresponding mirror map with the complex structure parameters. Even though the exact higher genus formulae around the orbifold point agree with the matrix model at weak t'Hooft coupling, its naive expansion around large t'Hooft coupling may still miss some contributions in the matrix model. For example, it is known that in quantum mechanics the same perturbative series can indeed give rise to distinct energy levels due to instanton effects \cite{ZinnJustin:2004}. More studies are needed to clarify this subtle issue. 

In \cite{Kallen:2014}, the author generalized the spectral problem of ABJM theory to the spectral problem of ABJ theory, and solve the spectrum by using the similar method as in \cite{Kallen:2013}. If the higher order smooth corrections to the non--perturbative contributions do exist, then the spectrum given in \cite{Kallen:2014} may also need to be revised.

\vspace{0.2in} {\leftline {\bf Acknowledgments}}
We thank Marcus Marino for correspondences regarding our previous paper \cite{Huang:201406}. This work is supported by the ``Young Thousand People" plan by the Central Organization Department in China, and by the Natural Science Foundation of China.

\clearpage
\begin{longtable}{| p{0.03\textwidth} | p{\textwidth} |} 
\hline $d$& $\sum_{d_1+d_2=d} \sum_{j_L,j_R} \oplus N^{d_1,d_2}_{j_L,j_R}(j_L,j_R)_{d_1,d_2} $ \\ \hline $1$ & ($0$,$\frac{1}{2}$)$_{0,1}$\\*[0.1cm] \hline $2$ & ($0$,$\frac{3}{2}$)$_{1,1}$\\*[0.1cm] \hline $3$ & ($0$,$\frac{5}{2}$)$_{1,2}$\\*[0.1cm] \hline $4$ & ($0$,$\frac{7}{2}$)$_{1,3}$$\oplus$($\frac{1}{2}$,$4$)$_{2,2}$$\oplus$($0$,$\frac{7}{2}$)$_{2,2}$$\oplus$($0$,$\frac{5}{2}$)$_{2,2}$\\*[0.1cm] \hline $5$ & ($0$,$\frac{9}{2}$)$_{1,4}$$\oplus$($1$,$\frac{11}{2}$)$_{2,3}$$\oplus$($\frac{1}{2}$,$5$)$_{2,3}$$\oplus$$2$($0$,$\frac{9}{2}$)$_{2,3}$$\oplus$($\frac{1}{2}$,$4$)$_{2,3}$$\oplus$($0$,$\frac{7}{2}$)$_{2,3}$$\oplus$($0$,$\frac{5}{2}$)$_{2,3}$\\*[0.1cm] \hline $6$ & ($0$,$\frac{11}{2}$)$_{1,5}$$\oplus$($\frac{3}{2}$,$7$)$_{2,4}$$\oplus$($1$,$\frac{13}{2}$)$_{2,4}$$\oplus$$2$($\frac{1}{2}$,$6$)$_{2,4}$$\oplus$($1$,$\frac{11}{2}$)$_{2,4}$$\oplus$$2$($0$,$\frac{11}{2}$)$_{2,4}$$\oplus$($\frac{1}{2}$,$5$)$_{2,4}$$\oplus$$2$($0$,$\frac{9}{2}$)$_{2,4}$$\oplus$($\frac{1}{2}$,$4$)$_{2,4}$\\*[0.1cm] & $\oplus$($0$,$\frac{7}{2}$)$_{2,4}$$\oplus$($0$,$\frac{5}{2}$)$_{2,4}$$\oplus$($2$,$\frac{15}{2}$)$_{3,3}$$\oplus$($\frac{3}{2}$,$7$)$_{3,3}$$\oplus$($\frac{1}{2}$,$7$)$_{3,3}$$\oplus$$3$($1$,$\frac{13}{2}$)$_{3,3}$$\oplus$($\frac{3}{2}$,$6$)$_{3,3}$$\oplus$$3$($\frac{1}{2}$,$6$)$_{3,3}$$\oplus$$2$($1$,$\frac{11}{2}$)$_{3,3}$\\*[0.1cm] & $\oplus$$4$($0$,$\frac{11}{2}$)$_{3,3}$$\oplus$$3$($\frac{1}{2}$,$5$)$_{3,3}$$\oplus$($1$,$\frac{9}{2}$)$_{3,3}$$\oplus$$3$($0$,$\frac{9}{2}$)$_{3,3}$$\oplus$$2$($\frac{1}{2}$,$4$)$_{3,3}$$\oplus$$3$($0$,$\frac{7}{2}$)$_{3,3}$$\oplus$($\frac{1}{2}$,$3$)$_{3,3}$$\oplus$($0$,$\frac{5}{2}$)$_{3,3}$$\oplus$($0$,$\frac{3}{2}$)$_{3,3}$\\*[0.1cm] \hline $7$ & ($0$,$\frac{13}{2}$)$_{1,6}$$\oplus$($2$,$\frac{17}{2}$)$_{2,5}$$\oplus$($\frac{3}{2}$,$8$)$_{2,5}$$\oplus$$2$($1$,$\frac{15}{2}$)$_{2,5}$$\oplus$($\frac{3}{2}$,$7$)$_{2,5}$$\oplus$$2$($\frac{1}{2}$,$7$)$_{2,5}$$\oplus$($1$,$\frac{13}{2}$)$_{2,5}$$\oplus$$3$($0$,$\frac{13}{2}$)$_{2,5}$$\oplus$$2$($\frac{1}{2}$,$6$)$_{2,5}$\\*[0.1cm] & $\oplus$($1$,$\frac{11}{2}$)$_{2,5}$$\oplus$$2$($0$,$\frac{11}{2}$)$_{2,5}$$\oplus$($\frac{1}{2}$,$5$)$_{2,5}$$\oplus$$2$($0$,$\frac{9}{2}$)$_{2,5}$$\oplus$($\frac{1}{2}$,$4$)$_{2,5}$$\oplus$($0$,$\frac{7}{2}$)$_{2,5}$$\oplus$($0$,$\frac{5}{2}$)$_{2,5}$$\oplus$($3$,$\frac{19}{2}$)$_{3,4}$$\oplus$($\frac{5}{2}$,$9$)$_{3,4}$\\*[0.1cm] & $\oplus$($\frac{3}{2}$,$9$)$_{3,4}$$\oplus$$3$($2$,$\frac{17}{2}$)$_{3,4}$$\oplus$($\frac{5}{2}$,$8$)$_{3,4}$$\oplus$($1$,$\frac{17}{2}$)$_{3,4}$$\oplus$$4$($\frac{3}{2}$,$8$)$_{3,4}$$\oplus$$2$($2$,$\frac{15}{2}$)$_{3,4}$$\oplus$($0$,$\frac{17}{2}$)$_{3,4}$$\oplus$$2$($\frac{1}{2}$,$8$)$_{3,4}$$\oplus$$7$($1$,$\frac{15}{2}$)$_{3,4}$\\*[0.1cm] & $\oplus$$4$($\frac{3}{2}$,$7$)$_{3,4}$$\oplus$($2$,$\frac{13}{2}$)$_{3,4}$$\oplus$($0$,$\frac{15}{2}$)$_{3,4}$$\oplus$$7$($\frac{1}{2}$,$7$)$_{3,4}$$\oplus$$6$($1$,$\frac{13}{2}$)$_{3,4}$$\oplus$$2$($\frac{3}{2}$,$6$)$_{3,4}$$\oplus$$7$($0$,$\frac{13}{2}$)$_{3,4}$$\oplus$$8$($\frac{1}{2}$,$6$)$_{3,4}$$\oplus$$5$($1$,$\frac{11}{2}$)$_{3,4}$\\*[0.1cm] & $\oplus$($\frac{3}{2}$,$5$)$_{3,4}$$\oplus$$6$($0$,$\frac{11}{2}$)$_{3,4}$$\oplus$$6$($\frac{1}{2}$,$5$)$_{3,4}$$\oplus$$2$($1$,$\frac{9}{2}$)$_{3,4}$$\oplus$$7$($0$,$\frac{9}{2}$)$_{3,4}$$\oplus$$4$($\frac{1}{2}$,$4$)$_{3,4}$$\oplus$($1$,$\frac{7}{2}$)$_{3,4}$$\oplus$$4$($0$,$\frac{7}{2}$)$_{3,4}$$\oplus$$2$($\frac{1}{2}$,$3$)$_{3,4}$\\*[0.1cm] & $\oplus$$3$($0$,$\frac{5}{2}$)$_{3,4}$$\oplus$($\frac{1}{2}$,$2$)$_{3,4}$$\oplus$($0$,$\frac{3}{2}$)$_{3,4}$$\oplus$($0$,$\frac{1}{2}$)$_{3,4}$\\*[0.1cm] \hline $8$ & ($0$,$\frac{15}{2}$)$_{1,7}$$\oplus$($\frac{5}{2}$,$10$)$_{2,6}$$\oplus$($2$,$\frac{19}{2}$)$_{2,6}$$\oplus$$2$($\frac{3}{2}$,$9$)$_{2,6}$$\oplus$($2$,$\frac{17}{2}$)$_{2,6}$$\oplus$$2$($1$,$\frac{17}{2}$)$_{2,6}$$\oplus$($\frac{3}{2}$,$8$)$_{2,6}$$\oplus$$3$($\frac{1}{2}$,$8$)$_{2,6}$\\*[0.1cm] & $\oplus$$2$($1$,$\frac{15}{2}$)$_{2,6}$$\oplus$($\frac{3}{2}$,$7$)$_{2,6}$$\oplus$$3$($0$,$\frac{15}{2}$)$_{2,6}$$\oplus$$2$($\frac{1}{2}$,$7$)$_{2,6}$$\oplus$($1$,$\frac{13}{2}$)$_{2,6}$$\oplus$$3$($0$,$\frac{13}{2}$)$_{2,6}$$\oplus$$2$($\frac{1}{2}$,$6$)$_{2,6}$$\oplus$($1$,$\frac{11}{2}$)$_{2,6}$$\oplus$$2$($0$,$\frac{11}{2}$)$_{2,6}$\\*[0.1cm] & $\oplus$($\frac{1}{2}$,$5$)$_{2,6}$$\oplus$$2$($0$,$\frac{9}{2}$)$_{2,6}$$\oplus$($\frac{1}{2}$,$4$)$_{2,6}$$\oplus$($0$,$\frac{7}{2}$)$_{2,6}$$\oplus$($0$,$\frac{5}{2}$)$_{2,6}$$\oplus$($4$,$\frac{23}{2}$)$_{3,5}$$\oplus$($\frac{7}{2}$,$11$)$_{3,5}$$\oplus$($\frac{5}{2}$,$11$)$_{3,5}$$\oplus$$3$($3$,$\frac{21}{2}$)$_{3,5}$\\*[0.1cm] & $\oplus$($\frac{7}{2}$,$10$)$_{3,5}$$\oplus$($2$,$\frac{21}{2}$)$_{3,5}$$\oplus$$4$($\frac{5}{2}$,$10$)$_{3,5}$$\oplus$$2$($3$,$\frac{19}{2}$)$_{3,5}$$\oplus$($1$,$\frac{21}{2}$)$_{3,5}$$\oplus$$3$($\frac{3}{2}$,$10$)$_{3,5}$$\oplus$$8$($2$,$\frac{19}{2}$)$_{3,5}$$\oplus$$4$($\frac{5}{2}$,$9$)$_{3,5}$\\*[0.1cm] & $\oplus$($3$,$\frac{17}{2}$)$_{3,5}$$\oplus$($\frac{1}{2}$,$10$)$_{3,5}$$\oplus$$3$($1$,$\frac{19}{2}$)$_{3,5}$$\oplus$$10$($\frac{3}{2}$,$9$)$_{3,5}$$\oplus$$7$($2$,$\frac{17}{2}$)$_{3,5}$$\oplus$$2$($\frac{5}{2}$,$8$)$_{3,5}$$\oplus$$2$($0$,$\frac{19}{2}$)$_{3,5}$$\oplus$$5$($\frac{1}{2}$,$9$)$_{3,5}$\\*[0.1cm] & $\oplus$$14$($1$,$\frac{17}{2}$)$_{3,5}$$\oplus$$11$($\frac{3}{2}$,$8$)$_{3,5}$$\oplus$$5$($2$,$\frac{15}{2}$)$_{3,5}$$\oplus$($\frac{5}{2}$,$7$)$_{3,5}$$\oplus$$3$($0$,$\frac{17}{2}$)$_{3,5}$$\oplus$$13$($\frac{1}{2}$,$8$)$_{3,5}$$\oplus$$14$($1$,$\frac{15}{2}$)$_{3,5}$$\oplus$$8$($\frac{3}{2}$,$7$)$_{3,5}$\\*[0.1cm] & $\oplus$$2$($2$,$\frac{13}{2}$)$_{3,5}$$\oplus$$12$($0$,$\frac{15}{2}$)$_{3,5}$$\oplus$$16$($\frac{1}{2}$,$7$)$_{3,5}$$\oplus$$13$($1$,$\frac{13}{2}$)$_{3,5}$$\oplus$$5$($\frac{3}{2}$,$6$)$_{3,5}$$\oplus$($2$,$\frac{11}{2}$)$_{3,5}$$\oplus$$11$($0$,$\frac{13}{2}$)$_{3,5}$$\oplus$$14$($\frac{1}{2}$,$6$)$_{3,5}$\\*[0.1cm] & $\oplus$$8$($1$,$\frac{11}{2}$)$_{3,5}$$\oplus$$2$($\frac{3}{2}$,$5$)$_{3,5}$$\oplus$$13$($0$,$\frac{11}{2}$)$_{3,5}$$\oplus$$11$($\frac{1}{2}$,$5$)$_{3,5}$$\oplus$$5$($1$,$\frac{9}{2}$)$_{3,5}$$\oplus$($\frac{3}{2}$,$4$)$_{3,5}$$\oplus$$9$($0$,$\frac{9}{2}$)$_{3,5}$$\oplus$$7$($\frac{1}{2}$,$4$)$_{3,5}$$\oplus$$2$($1$,$\frac{7}{2}$)$_{3,5}$\\*[0.1cm] & $\oplus$$8$($0$,$\frac{7}{2}$)$_{3,5}$$\oplus$$4$($\frac{1}{2}$,$3$)$_{3,5}$$\oplus$($1$,$\frac{5}{2}$)$_{3,5}$$\oplus$$4$($0$,$\frac{5}{2}$)$_{3,5}$$\oplus$$2$($\frac{1}{2}$,$2$)$_{3,5}$$\oplus$$3$($0$,$\frac{3}{2}$)$_{3,5}$$\oplus$($\frac{1}{2}$,$1$)$_{3,5}$$\oplus$($0$,$\frac{1}{2}$)$_{3,5}$$\oplus$($\frac{9}{2}$,$12$)$_{4,4}$\\*[0.1cm] & $\oplus$($4$,$\frac{23}{2}$)$_{4,4}$$\oplus$($3$,$\frac{23}{2}$)$_{4,4}$$\oplus$$3$($\frac{7}{2}$,$11$)$_{4,4}$$\oplus$($4$,$\frac{21}{2}$)$_{4,4}$$\oplus$$2$($\frac{5}{2}$,$11$)$_{4,4}$$\oplus$$5$($3$,$\frac{21}{2}$)$_{4,4}$$\oplus$$2$($\frac{7}{2}$,$10$)$_{4,4}$$\oplus$($\frac{3}{2}$,$11$)$_{4,4}$\\*[0.1cm] & $\oplus$$3$($2$,$\frac{21}{2}$)$_{4,4}$$\oplus$$9$($\frac{5}{2}$,$10$)$_{4,4}$$\oplus$$5$($3$,$\frac{19}{2}$)$_{4,4}$$\oplus$($\frac{7}{2}$,$9$)$_{4,4}$$\oplus$$2$($1$,$\frac{21}{2}$)$_{4,4}$$\oplus$$6$($\frac{3}{2}$,$10$)$_{4,4}$$\oplus$$13$($2$,$\frac{19}{2}$)$_{4,4}$$\oplus$$8$($\frac{5}{2}$,$9$)$_{4,4}$\\*[0.1cm] & $\oplus$$2$($3$,$\frac{17}{2}$)$_{4,4}$$\oplus$$3$($\frac{1}{2}$,$10$)$_{4,4}$$\oplus$$8$($1$,$\frac{19}{2}$)$_{4,4}$$\oplus$$20$($\frac{3}{2}$,$9$)$_{4,4}$$\oplus$$15$($2$,$\frac{17}{2}$)$_{4,4}$$\oplus$$6$($\frac{5}{2}$,$8$)$_{4,4}$$\oplus$($3$,$\frac{15}{2}$)$_{4,4}$$\oplus$$4$($0$,$\frac{19}{2}$)$_{4,4}$\\*[0.1cm] & $\oplus$$10$($\frac{1}{2}$,$9$)$_{4,4}$$\oplus$$23$($1$,$\frac{17}{2}$)$_{4,4}$$\oplus$$20$($\frac{3}{2}$,$8$)$_{4,4}$$\oplus$$10$($2$,$\frac{15}{2}$)$_{4,4}$$\oplus$$2$($\frac{5}{2}$,$7$)$_{4,4}$$\oplus$$8$($0$,$\frac{17}{2}$)$_{4,4}$$\oplus$$25$($\frac{1}{2}$,$8$)$_{4,4}$$\oplus$$28$($1$,$\frac{15}{2}$)$_{4,4}$\\*[0.1cm] & $\oplus$$18$($\frac{3}{2}$,$7$)$_{4,4}$$\oplus$$6$($2$,$\frac{13}{2}$)$_{4,4}$$\oplus$($\frac{5}{2}$,$6$)$_{4,4}$$\oplus$$18$($0$,$\frac{15}{2}$)$_{4,4}$$\oplus$$28$($\frac{1}{2}$,$7$)$_{4,4}$$\oplus$$24$($1$,$\frac{13}{2}$)$_{4,4}$$\oplus$$10$($\frac{3}{2}$,$6$)$_{4,4}$$\oplus$$2$($2$,$\frac{11}{2}$)$_{4,4}$\\*[0.1cm] & $\oplus$$21$($0$,$\frac{13}{2}$)$_{4,4}$$\oplus$$28$($\frac{1}{2}$,$6$)$_{4,4}$$\oplus$$18$($1$,$\frac{11}{2}$)$_{4,4}$$\oplus$$6$($\frac{3}{2}$,$5$)$_{4,4}$$\oplus$($2$,$\frac{9}{2}$)$_{4,4}$$\oplus$$20$($0$,$\frac{11}{2}$)$_{4,4}$$\oplus$$20$($\frac{1}{2}$,$5$)$_{4,4}$$\oplus$$10$($1$,$\frac{9}{2}$)$_{4,4}$\\*[0.1cm] & $\oplus$$2$($\frac{3}{2}$,$4$)$_{4,4}$$\oplus$$18$($0$,$\frac{9}{2}$)$_{4,4}$$\oplus$$15$($\frac{1}{2}$,$4$)$_{4,4}$$\oplus$$6$($1$,$\frac{7}{2}$)$_{4,4}$$\oplus$($\frac{3}{2}$,$3$)$_{4,4}$$\oplus$$12$($0$,$\frac{7}{2}$)$_{4,4}$$\oplus$$8$($\frac{1}{2}$,$3$)$_{4,4}$$\oplus$$2$($1$,$\frac{5}{2}$)$_{4,4}$$\oplus$$9$($0$,$\frac{5}{2}$)$_{4,4}$\\*[0.1cm] & $\oplus$$5$($\frac{1}{2}$,$2$)$_{4,4}$$\oplus$($1$,$\frac{3}{2}$)$_{4,4}$$\oplus$$5$($0$,$\frac{3}{2}$)$_{4,4}$$\oplus$$2$($\frac{1}{2}$,$1$)$_{4,4}$$\oplus$$3$($0$,$\frac{1}{2}$)$_{4,4}$\\*[0.1cm] \hline $9$ & ($0$,$\frac{17}{2}$)$_{1,8}$$\oplus$($3$,$\frac{23}{2}$)$_{2,7}$$\oplus$($\frac{5}{2}$,$11$)$_{2,7}$$\oplus$$2$($2$,$\frac{21}{2}$)$_{2,7}$$\oplus$($\frac{5}{2}$,$10$)$_{2,7}$$\oplus$$2$($\frac{3}{2}$,$10$)$_{2,7}$$\oplus$($2$,$\frac{19}{2}$)$_{2,7}$$\oplus$$3$($1$,$\frac{19}{2}$)$_{2,7}$\\*[0.1cm] & $\oplus$$2$($\frac{3}{2}$,$9$)$_{2,7}$$\oplus$($2$,$\frac{17}{2}$)$_{2,7}$$\oplus$$3$($\frac{1}{2}$,$9$)$_{2,7}$$\oplus$$2$($1$,$\frac{17}{2}$)$_{2,7}$$\oplus$($\frac{3}{2}$,$8$)$_{2,7}$$\oplus$$4$($0$,$\frac{17}{2}$)$_{2,7}$$\oplus$$3$($\frac{1}{2}$,$8$)$_{2,7}$$\oplus$$2$($1$,$\frac{15}{2}$)$_{2,7}$$\oplus$($\frac{3}{2}$,$7$)$_{2,7}$\\*[0.1cm] & $\oplus$$3$($0$,$\frac{15}{2}$)$_{2,7}$$\oplus$$2$($\frac{1}{2}$,$7$)$_{2,7}$$\oplus$($1$,$\frac{13}{2}$)$_{2,7}$$\oplus$$3$($0$,$\frac{13}{2}$)$_{2,7}$$\oplus$$2$($\frac{1}{2}$,$6$)$_{2,7}$$\oplus$($1$,$\frac{11}{2}$)$_{2,7}$$\oplus$$2$($0$,$\frac{11}{2}$)$_{2,7}$$\oplus$($\frac{1}{2}$,$5$)$_{2,7}$$\oplus$$2$($0$,$\frac{9}{2}$)$_{2,7}$\\*[0.1cm] & $\oplus$($\frac{1}{2}$,$4$)$_{2,7}$$\oplus$($0$,$\frac{7}{2}$)$_{2,7}$$\oplus$($0$,$\frac{5}{2}$)$_{2,7}$$\oplus$($5$,$\frac{27}{2}$)$_{3,6}$$\oplus$($\frac{9}{2}$,$13$)$_{3,6}$$\oplus$($\frac{7}{2}$,$13$)$_{3,6}$$\oplus$$3$($4$,$\frac{25}{2}$)$_{3,6}$$\oplus$($\frac{9}{2}$,$12$)$_{3,6}$$\oplus$($3$,$\frac{25}{2}$)$_{3,6}$\\*[0.1cm] & $\oplus$$4$($\frac{7}{2}$,$12$)$_{3,6}$$\oplus$$2$($4$,$\frac{23}{2}$)$_{3,6}$$\oplus$($2$,$\frac{25}{2}$)$_{3,6}$$\oplus$$3$($\frac{5}{2}$,$12$)$_{3,6}$$\oplus$$8$($3$,$\frac{23}{2}$)$_{3,6}$$\oplus$$4$($\frac{7}{2}$,$11$)$_{3,6}$$\oplus$($4$,$\frac{21}{2}$)$_{3,6}$$\oplus$($\frac{3}{2}$,$12$)$_{3,6}$\\*[0.1cm] & $\oplus$$4$($2$,$\frac{23}{2}$)$_{3,6}$$\oplus$$11$($\frac{5}{2}$,$11$)$_{3,6}$$\oplus$$7$($3$,$\frac{21}{2}$)$_{3,6}$$\oplus$$2$($\frac{7}{2}$,$10$)$_{3,6}$$\oplus$($\frac{1}{2}$,$12$)$_{3,6}$$\oplus$$3$($1$,$\frac{23}{2}$)$_{3,6}$$\oplus$$7$($\frac{3}{2}$,$11$)$_{3,6}$$\oplus$$17$($2$,$\frac{21}{2}$)$_{3,6}$\\*[0.1cm] & $\oplus$$12$($\frac{5}{2}$,$10$)$_{3,6}$$\oplus$$5$($3$,$\frac{19}{2}$)$_{3,6}$$\oplus$($\frac{7}{2}$,$9$)$_{3,6}$$\oplus$$3$($\frac{1}{2}$,$11$)$_{3,6}$$\oplus$$8$($1$,$\frac{21}{2}$)$_{3,6}$$\oplus$$20$($\frac{3}{2}$,$10$)$_{3,6}$$\oplus$$17$($2$,$\frac{19}{2}$)$_{3,6}$$\oplus$$8$($\frac{5}{2}$,$9$)$_{3,6}$\\*[0.1cm] & $\oplus$$2$($3$,$\frac{17}{2}$)$_{3,6}$$\oplus$$4$($0$,$\frac{21}{2}$)$_{3,6}$$\oplus$$10$($\frac{1}{2}$,$10$)$_{3,6}$$\oplus$$25$($1$,$\frac{19}{2}$)$_{3,6}$$\oplus$$24$($\frac{3}{2}$,$9$)$_{3,6}$$\oplus$$15$($2$,$\frac{17}{2}$)$_{3,6}$$\oplus$$5$($\frac{5}{2}$,$8$)$_{3,6}$$\oplus$($3$,$\frac{15}{2}$)$_{3,6}$\\*[0.1cm] & $\oplus$$6$($0$,$\frac{19}{2}$)$_{3,6}$$\oplus$$23$($\frac{1}{2}$,$9$)$_{3,6}$$\oplus$$27$($1$,$\frac{17}{2}$)$_{3,6}$$\oplus$$20$($\frac{3}{2}$,$8$)$_{3,6}$$\oplus$$9$($2$,$\frac{15}{2}$)$_{3,6}$$\oplus$$2$($\frac{5}{2}$,$7$)$_{3,6}$$\oplus$$19$($0$,$\frac{17}{2}$)$_{3,6}$$\oplus$$28$($\frac{1}{2}$,$8$)$_{3,6}$\\*[0.1cm] \hline $9$ & $\oplus$$27$($1$,$\frac{15}{2}$)$_{3,6}$$\oplus$$15$($\frac{3}{2}$,$7$)$_{3,6}$$\oplus$$5$($2$,$\frac{13}{2}$)$_{3,6}$$\oplus$($\frac{5}{2}$,$6$)$_{3,6}$$\oplus$$19$($0$,$\frac{15}{2}$)$_{3,6}$$\oplus$$27$($\frac{1}{2}$,$7$)$_{3,6}$$\oplus$$20$($1$,$\frac{13}{2}$)$_{3,6}$$\oplus$$9$($\frac{3}{2}$,$6$)$_{3,6}$\\*[0.1cm] & $\oplus$$2$($2$,$\frac{11}{2}$)$_{3,6}$$\oplus$$22$($0$,$\frac{13}{2}$)$_{3,6}$$\oplus$$23$($\frac{1}{2}$,$6$)$_{3,6}$$\oplus$$15$($1$,$\frac{11}{2}$)$_{3,6}$$\oplus$$5$($\frac{3}{2}$,$5$)$_{3,6}$$\oplus$($2$,$\frac{9}{2}$)$_{3,6}$$\oplus$$17$($0$,$\frac{11}{2}$)$_{3,6}$$\oplus$$17$($\frac{1}{2}$,$5$)$_{3,6}$\\*[0.1cm] & $\oplus$$8$($1$,$\frac{9}{2}$)$_{3,6}$$\oplus$$2$($\frac{3}{2}$,$4$)$_{3,6}$$\oplus$$16$($0$,$\frac{9}{2}$)$_{3,6}$$\oplus$$12$($\frac{1}{2}$,$4$)$_{3,6}$$\oplus$$5$($1$,$\frac{7}{2}$)$_{3,6}$$\oplus$($\frac{3}{2}$,$3$)$_{3,6}$$\oplus$$10$($0$,$\frac{7}{2}$)$_{3,6}$$\oplus$$7$($\frac{1}{2}$,$3$)$_{3,6}$$\oplus$$2$($1$,$\frac{5}{2}$)$_{3,6}$\\*[0.1cm] & $\oplus$$8$($0$,$\frac{5}{2}$)$_{3,6}$$\oplus$$4$($\frac{1}{2}$,$2$)$_{3,6}$$\oplus$($1$,$\frac{3}{2}$)$_{3,6}$$\oplus$$4$($0$,$\frac{3}{2}$)$_{3,6}$$\oplus$$2$($\frac{1}{2}$,$1$)$_{3,6}$$\oplus$$2$($0$,$\frac{1}{2}$)$_{3,6}$$\oplus$($\frac{1}{2}$,$0$)$_{3,6}$$\oplus$($6$,$\frac{29}{2}$)$_{4,5}$$\oplus$($\frac{11}{2}$,$14$)$_{4,5}$\\*[0.1cm] & $\oplus$($\frac{9}{2}$,$14$)$_{4,5}$$\oplus$$3$($5$,$\frac{27}{2}$)$_{4,5}$$\oplus$($\frac{11}{2}$,$13$)$_{4,5}$$\oplus$$2$($4$,$\frac{27}{2}$)$_{4,5}$$\oplus$$5$($\frac{9}{2}$,$13$)$_{4,5}$$\oplus$$2$($5$,$\frac{25}{2}$)$_{4,5}$$\oplus$($3$,$\frac{27}{2}$)$_{4,5}$$\oplus$$4$($\frac{7}{2}$,$13$)$_{4,5}$\\*[0.1cm] & $\oplus$$10$($4$,$\frac{25}{2}$)$_{4,5}$$\oplus$$5$($\frac{9}{2}$,$12$)$_{4,5}$$\oplus$($5$,$\frac{23}{2}$)$_{4,5}$$\oplus$$2$($\frac{5}{2}$,$13$)$_{4,5}$$\oplus$$7$($3$,$\frac{25}{2}$)$_{4,5}$$\oplus$$15$($\frac{7}{2}$,$12$)$_{4,5}$$\oplus$$9$($4$,$\frac{23}{2}$)$_{4,5}$$\oplus$$2$($\frac{9}{2}$,$11$)$_{4,5}$\\*[0.1cm] & $\oplus$($\frac{3}{2}$,$13$)$_{4,5}$$\oplus$$5$($2$,$\frac{25}{2}$)$_{4,5}$$\oplus$$13$($\frac{5}{2}$,$12$)$_{4,5}$$\oplus$$26$($3$,$\frac{23}{2}$)$_{4,5}$$\oplus$$17$($\frac{7}{2}$,$11$)$_{4,5}$$\oplus$$6$($4$,$\frac{21}{2}$)$_{4,5}$$\oplus$($\frac{9}{2}$,$10$)$_{4,5}$$\oplus$($1$,$\frac{25}{2}$)$_{4,5}$\\*[0.1cm] & $\oplus$$7$($\frac{3}{2}$,$12$)$_{4,5}$$\oplus$$18$($2$,$\frac{23}{2}$)$_{4,5}$$\oplus$$35$($\frac{5}{2}$,$11$)$_{4,5}$$\oplus$$27$($3$,$\frac{21}{2}$)$_{4,5}$$\oplus$$11$($\frac{7}{2}$,$10$)$_{4,5}$$\oplus$$2$($4$,$\frac{19}{2}$)$_{4,5}$$\oplus$($0$,$\frac{25}{2}$)$_{4,5}$$\oplus$$3$($\frac{1}{2}$,$12$)$_{4,5}$\\*[0.1cm] & $\oplus$$13$($1$,$\frac{23}{2}$)$_{4,5}$$\oplus$$28$($\frac{3}{2}$,$11$)$_{4,5}$$\oplus$$51$($2$,$\frac{21}{2}$)$_{4,5}$$\oplus$$42$($\frac{5}{2}$,$10$)$_{4,5}$$\oplus$$22$($3$,$\frac{19}{2}$)$_{4,5}$$\oplus$$6$($\frac{7}{2}$,$9$)$_{4,5}$$\oplus$($4$,$\frac{17}{2}$)$_{4,5}$$\oplus$$2$($0$,$\frac{23}{2}$)$_{4,5}$\\*[0.1cm] & $\oplus$$14$($\frac{1}{2}$,$11$)$_{4,5}$$\oplus$$33$($1$,$\frac{21}{2}$)$_{4,5}$$\oplus$$63$($\frac{3}{2}$,$10$)$_{4,5}$$\oplus$$58$($2$,$\frac{19}{2}$)$_{4,5}$$\oplus$$34$($\frac{5}{2}$,$9$)$_{4,5}$$\oplus$$12$($3$,$\frac{17}{2}$)$_{4,5}$$\oplus$$2$($\frac{7}{2}$,$8$)$_{4,5}$$\oplus$$15$($0$,$\frac{21}{2}$)$_{4,5}$\\*[0.1cm] & $\oplus$$37$($\frac{1}{2}$,$10$)$_{4,5}$$\oplus$$74$($1$,$\frac{19}{2}$)$_{4,5}$$\oplus$$77$($\frac{3}{2}$,$9$)$_{4,5}$$\oplus$$54$($2$,$\frac{17}{2}$)$_{4,5}$$\oplus$$23$($\frac{5}{2}$,$8$)$_{4,5}$$\oplus$$6$($3$,$\frac{15}{2}$)$_{4,5}$$\oplus$($\frac{7}{2}$,$7$)$_{4,5}$$\oplus$$25$($0$,$\frac{19}{2}$)$_{4,5}$\\*[0.1cm] & $\oplus$$69$($\frac{1}{2}$,$9$)$_{4,5}$$\oplus$$87$($1$,$\frac{17}{2}$)$_{4,5}$$\oplus$$71$($\frac{3}{2}$,$8$)$_{4,5}$$\oplus$$37$($2$,$\frac{15}{2}$)$_{4,5}$$\oplus$$12$($\frac{5}{2}$,$7$)$_{4,5}$$\oplus$$2$($3$,$\frac{13}{2}$)$_{4,5}$$\oplus$$51$($0$,$\frac{17}{2}$)$_{4,5}$$\oplus$$85$($\frac{1}{2}$,$8$)$_{4,5}$\\*[0.1cm] & $\oplus$$89$($1$,$\frac{15}{2}$)$_{4,5}$$\oplus$$56$($\frac{3}{2}$,$7$)$_{4,5}$$\oplus$$24$($2$,$\frac{13}{2}$)$_{4,5}$$\oplus$$6$($\frac{5}{2}$,$6$)$_{4,5}$$\oplus$($3$,$\frac{11}{2}$)$_{4,5}$$\oplus$$54$($0$,$\frac{15}{2}$)$_{4,5}$$\oplus$$85$($\frac{1}{2}$,$7$)$_{4,5}$$\oplus$$70$($1$,$\frac{13}{2}$)$_{4,5}$\\*[0.1cm] & $\oplus$$37$($\frac{3}{2}$,$6$)$_{4,5}$$\oplus$$12$($2$,$\frac{11}{2}$)$_{4,5}$$\oplus$$2$($\frac{5}{2}$,$5$)$_{4,5}$$\oplus$$60$($0$,$\frac{13}{2}$)$_{4,5}$$\oplus$$74$($\frac{1}{2}$,$6$)$_{4,5}$$\oplus$$54$($1$,$\frac{11}{2}$)$_{4,5}$$\oplus$$23$($\frac{3}{2}$,$5$)$_{4,5}$$\oplus$$6$($2$,$\frac{9}{2}$)$_{4,5}$\\*[0.1cm] & $\oplus$($\frac{5}{2}$,$4$)$_{4,5}$$\oplus$$50$($0$,$\frac{11}{2}$)$_{4,5}$$\oplus$$57$($\frac{1}{2}$,$5$)$_{4,5}$$\oplus$$34$($1$,$\frac{9}{2}$)$_{4,5}$$\oplus$$12$($\frac{3}{2}$,$4$)$_{4,5}$$\oplus$$2$($2$,$\frac{7}{2}$)$_{4,5}$$\oplus$$44$($0$,$\frac{9}{2}$)$_{4,5}$$\oplus$$41$($\frac{1}{2}$,$4$)$_{4,5}$\\*[0.1cm] & $\oplus$$22$($1$,$\frac{7}{2}$)$_{4,5}$$\oplus$$6$($\frac{3}{2}$,$3$)$_{4,5}$$\oplus$($2$,$\frac{5}{2}$)$_{4,5}$$\oplus$$30$($0$,$\frac{7}{2}$)$_{4,5}$$\oplus$$27$($\frac{1}{2}$,$3$)$_{4,5}$$\oplus$$11$($1$,$\frac{5}{2}$)$_{4,5}$$\oplus$$2$($\frac{3}{2}$,$2$)$_{4,5}$$\oplus$$24$($0$,$\frac{5}{2}$)$_{4,5}$$\oplus$$17$($\frac{1}{2}$,$2$)$_{4,5}$\\*[0.1cm] & $\oplus$$5$($1$,$\frac{3}{2}$)$_{4,5}$$\oplus$($\frac{3}{2}$,$1$)$_{4,5}$$\oplus$$14$($0$,$\frac{3}{2}$)$_{4,5}$$\oplus$$8$($\frac{1}{2}$,$1$)$_{4,5}$$\oplus$$2$($1$,$\frac{1}{2}$)$_{4,5}$$\oplus$$8$($0$,$\frac{1}{2}$)$_{4,5}$$\oplus$$3$($\frac{1}{2}$,$0$)$_{4,5}$\\*[0.1cm] \hline
\caption{The GV invariants $N_{j_L,j_R}^{d_1,d_2}$ for $d=1,2, \cdots ,9$ for the local $\mathbb{P}^1\times \mathbb{P}^1$ model. Here $d_1,d_2$ denote the degrees of the base $\mathbb{P}^1$ and the fiber $\mathbb{P}^1$. There is a symmetry $N_{j_L,j_R}^{d_1,d_2}=N_{j_L,j_R}^{d_2,d_1}$ since the fibration is trivial. So we only list the cases $d_1\geqslant d_2$.}\label{GVtable}
\end{longtable}

\end{document}